# Gold assisted molecular beam epitaxy of Ge nanostructures on Ge (100) Surface

A. Rath[1], J. K. Dash[1], R. R. Juluri[1], A. Ghosh[1] and P. V. Satyam[1,*]

[1] Institute of Physics, Sachivalaya Marg, Bhubaneswar - 751005, India

**Abstract**

We report on the gold assisted epitaxial growth of Ge nanostructures under ultra high vacuum (UHV) conditions ($\approx 3 \times 10^{-10}$ mbar) on clean Ge (100) surfaces. For this study, $\approx 2.0$ nm thick Au samples were grown on the substrate surface by molecular beam epitaxy (MBE). Thermal annealing was carried out inside the UHV chamber at temperature $\approx 500\,°C$ and following this, well ordered gold nanostructures placed on pedestal Ge were formed. A $\approx 2$ nm Ge film was further deposited on the above surface while the substrate was kept at a temperature of $\approx 500\,°C$. The height of the Ge (pedestal) underneath gold increased along with the formation of small Ge islands. The effect of substrate temperature and role of gold on the formation of above structures has been discussed in detail. Electron microscopy (TEM, SEM) studies were carried out to determine the structure of Au – Ge nano systems.

## 1. Introduction

Epitaxy of thin films is a key process in modern nano electronic industries. It allows abrupt doping profiles in homoepitaxy and the formation of sharp interfaces in heteroepitaxy to be realized. More than 90 % of semiconductor devices consist of silicon. Knowledge of surface properties of Si, Ge and Au nanostructures are crucial for rapidly developing area of nanotechnology. Successful implementation of germanium technology will however require an understanding of the solid-state interaction in metal-germanium systems [1-3]. As the feature size becomes smaller the transistor density increases and with it the heat production. This is a problem that can be compensated by using lower driving voltages. To allow this, a higher electron and hole mobility, relating the drift current to the applied electric field, is required from the semiconductor. For this, Ge is the right material to achieve this goal because it has high mobility for both electrons and holes [4]. Also due to the low vapor pressure, Au-Ge alloys are used as catalyst for the growth of Ge nano wires inside the UHV chamber [5].

The effect of size on phase stability and phase transformations is of both fundamental and applied interest. For example, during the nucleation and growth of self-assembled nanowires from nanoscale metal catalysts, the phase of the catalyst determines properties such as the growth rate and the structure of the nanowire. Any size-dependent or growth-rate dependent changes in the catalyst may thus have strong effects on the structures that formed [6–11]. In last seven years, interesting works have been carried out in understanding growth of silicon and germanium nanowire in presence of Au as catalyst and deviation in phase diagram of Au-Si and Au-Ge systems by using in-situ TEM methods [6–11]. Epitaxially grown self organized nanostructures on silicon have been studied in great detail [12-15]. Several works has also been done using the Ge substrate instead of Si [16]. Data concerning the behavior of metal thin films on germanium upon thermal treatment is relatively scarce. The thin film reactions of 20 transition metals, excluding gold, with germanium substrates have been reported by Gaudet et al [17].

The main focus of this work is on solid source molecular beam epitaxy of both Au and Ge on Ge (100) substrate. It is similar to the vapor-liquid-solid (VLS) growth of Ge with Au as transport medium, except molecular-beam source is used instead of a chemical compound vapor like CVD growth. This avoids many problems, including the high temperature required to decompose the vapor compound and possible contamination of the growing layer. Also, the deposition will be atom by atom which ensures uniform growth of the metal and the semiconductor material at the first stage of the process. Hence, adatom deposition is not selective and growth is driven by adatom diffusion on the substrate.

The solid solubility of Ge in Au is 3.1 %, but that of Au in Ge is very low ($<10^{-5}$ %) [18]. Systematic results reported by Sutter et al. clearly point out the deviation of Ge solubility from bulk for nano-scale systems (typically less than 50 nm size) [19]. The systems dealt by Sutter et al., are basically for VLS type growth (free standing out-of-plane NW growth) and the results showed that the phase diagram would deviate from the bulk [19]. Cheuh *et al* reported on the post growth engineering of the nanowire (NW) structures and composition through the alloying and phase segregation that is induced by thermal annealing in Au-Ge system [20]. A direct observation of the VLS growth of Ge nanowires was reported by Wu and Yang [21], who identified the various growth stages in correlation to the Au–Ge binary phase diagram. In our earlier work, we demonstrated that, gold-silicide nano alloy formation at the substrate (Si) surface is necessary for forming phase separated Au-Ge bilobed nanostructures. Our results also indicate that Si-Ge bonding is more favorable to have than having Au-Ge bond. It would be very interesting to study the structural changes in the above Au-Ge nanostructures, when Si substrate will be replaced by Ge substrate under the same ambient [22].

In this article, we present the experimental observation of growth of Au-Ge nano-structures during *in-situ* thermal treatment inside UHV chamber for gold deposited on Ge(100) surfaces without oxide layer at the interface followed by further deposition of Ge on it. E*x-situ* electron microscopy measurements (both SEM and TEM) confirm the presence of Au and Ge in these structures. The formation of Au-Ge nanostructures via nanoscale phase separation is demonstrated experimentally. Also, the role of substrate temperature in formation of these structures has been clearly demonstrated in this work.

## 2. Experimental

For this present study, we have prepared five types of samples. For the first case (sample A), a thin Au film of thickness of about ≈2.0 nm on n-type Ge(100), by MBE method under ultra high vacuum conditions [23]. Ge (100) substrates were loaded into the MBE chamber and degassed at ≈ 400°C for about 12 hours inside the chamber and followed by flashing for about 1 minute by direct heating at a temperature of ≈800°C. In this process, native oxide was removed and a clean Ge(100) surface was obtained. On such ultra clean surfaces, ≈2.0 nm thick gold films were grown epitaxially by evaporating Au from a Knudsen cell. Deposition rate was kept constant at ≈ 0.14 nm min$^{-1}$. Thermal annealing of the as-deposited sample (sample A) was carried out inside the UHV chamber at temperature of 500°C with a ramp rate of 7°C min$^{-1}$(sample B). Also, we deposited a ≈ 2.0 nm gold on Ge(100) sample in the MBE chamber as explained above (like sample A) and about 2.0 nm Ge was further deposited at a typical deposition rate of 0.6 ML/min (where one ML corresponds to 6.78 × 10$^{14}$ atoms cm$^{-2}$ for a Si(100) surface) with various substrate temperatures: 500°C (sample C) and 600°C(sample D). It should be noted that Au deposition and annealing and Ge deposition after Au deposition are done in sequentially without any break in the vacuum. In addition to this, we deposited 2.0 nm Ge on Ge((100) substrate followed by annealing upto 500°C (sample E). During the growth, chamber vacuum was ≈ 6.2 ×10$^{-10}$ mbar. The post-growth characterization of the above samples were done with the field emission gun based scanning electron microscopy (FEGSEM) measurements with 20 kV electrons using a Neon 40 cross-beam system (M/S Carl Zeiss GmbH). Cross-sectional TEM specimens were prepared from the above samples in which electron transparency was achieved through low energy Ar$^+$ ion milling. TEM measurements were performed with 200 keV electrons (2010, JEOL HRTEM) under cross-sectional geometries.

**3. Results and discussions**

Fig. 1(a) depicts a SEM micrograph of as deposited 2.0 nm Au Film on Ge(100) using MBE (sample A) with surface coverage of ~21%. Irregular and isolated gold nanostructures of typical size ~ 27 nm were formed. Fig.1 (b) shows the bright field XTEM image of the interface of gold film and the Ge substrate and corresponding high resolution image(HR) shows the d-spacing of Au(200) and Ge(111) plane (inset Fig.). The as-deposited sample was annealed *in-situ* inside the UHV chamber at 500°C for 30 minutes (sample B). Following this annealing in UHV, ex-situ SEM measurements (at room temperature) showed well distributed gold nanostructures placed on pedestal Ge (Fig. 2 (a)). Fig. 2(b) and 2(c) depicts the XTEM image of the Au nanostructures and cross-sectional HRTEM image of single Au nanostructures respectively. The contrast shows the inter diffusion of gold into the Ge substrate. In our earlier work [22], we had shown that for Au-Si system, gold nano rectangles were formed without having pedestal Si. The reason behind the formation is still not well understood.

When a ≈ 2 nm Ge was further deposited (sample C) using MBE system on the above Au-patterned surface at substrate temperature 500°C inside the UHV chamber, Au nanostructures with more prominent Ge pedestal were formed (Fig. 3 (a)). This means that Au-Ge nanalloys are acted as seeding positions for the growth of above structures. Because of the incorporation of larger amount of Ge in to the seed particle, height of the pedestal increases. Kodambaka et al. have shown using in situ microscopy for the Ge/Au system that catalysts can be either liquid or solid below eutectic temperature, depending on thermal history. They have shown that with a pre-annealing at 400 °C, the Ge/Au catalyst remains liquid down to 255°C [24], this phenomenon has also been reported very recently by Gamalski et al. [25]. This means that with our growth conditions the catalyst should be still liquid. Sample was then exposed to molecular beam of Ge. In Fig. 3(a), SEM image taken at 54° tilt clearly shows the three dimensional nature of the Au-Ge nanostructures, where gold is on the top of the Ge. The bright contrast is the gold and other one is for the Ge. As Au is higher Z material, more secondary electrons are emitted from the Au region than from Ge region and this causes the contrast difference between gold and germanium in SEM images. In addition to these structures, one can see the formation of Ge nanostructures having typical height ~ 5.0 nm (Fig. 3(b)). In Fig. 3(b), bright field XTEM images of the Au-Ge nanostructures along with Ge nano islands has been shown and corresponding high resolution XTEM of the epitaxially grown Ge nano structures is shown in the inset. Fig. 3(c) depicted HRXTEM micrograph of one of the Au-Ge structures. The typical height of the pedestal Ge is ~ 8.0 nm. Furthermore, one can see that the height of the pedestal Ge is several times larger than the effective thickness of the deposited Ge. The filtered image of the square marked region is demonstrated in Fig. 3(d). It shows the formation of Au-Ge composition at the interface of top gold and pedestal Ge. The d-spacing has been found to be of about 0.332 nm which do not match with the Au or Ge but matches quite well with the (224) plane of tetragonal phase of $Au_{0.6}Ge_{0.4}$ [26].

To do the ex-situ temperature dependent study, we deposited 2.0 nm Ge on Au patterned surface at 600°C (sample D) instead of 500°C. Formation of the above Au-Ge structures having typical size of the pedestal ~ 20 nm has been observed (Fig. 4(a)). It should be noted that, there is no formation of small Ge islands in this case and the height of the pedestal is more than 500°C case. In Fig. 4(b), one can clearly see the only formation of Au-Ge nanostructures. At higher enough temperature (600°C), atoms are arriving into the seeding particle not only from the molecular beam but also from the Ge islands due to self diffusion. This resulted in increase in height of the pedestal Ge. Fig. 4(c) depicted the HRXTEM micrograph of one of such structures. The filtered image of the squared marked region is shown in Fig. 4(d). The d-spacing shows the formation of $Au_{0.6}Ge_{0.4}$ [26] composition at the junction of top gold and pedestal Ge. In both the cases, the Au-Ge composition was found at the interface. It has been studied that the composition is not only depend on the annealing temperature but also on the cooling rate [16]. Thus, there will be some composition gradient at the interface.

From the above studies, it was found that the heated Au-Ge alloy structures may play a major role in formation of such Au-Ge structures. During the deposition of Ge, gold germanide particles act as nucleation center for the growth of Ge. As the deposition continues, the smaller liquid droplets provide a rapid diffusion path for the incoming Ge atoms, which find it energetically favorable to join already-nucleated crystals rather than to nucleate new ones. As substrate is also Ge and the Ge-Ge bonding is more favorable than Au-Ge bonding, Ge pedestal were formed underneath gold. Another interesting aspect of this study is the formation of Ge islands. In sample C, well distributed Ge nanoislands were formed along with Au-Ge structures. To study the effect of gold over layers on formation of islands, 2 nm Ge was deposited on cleaned Ge (100) surface followed by annealing at 500°C (sample E). Formation of Ge nanoislands were not seen (Fig. 5). In Fig. 5, the HRXTEM shows the homo epitaxial growth of Ge layer on the Ge(100) surface without islanding. The dotted white line is the sharp interface between the Ge layer and the Ge substrate. Thus presence of gold layer resulted in formation of Ge islands.

Numerous interesting aspects of Au-Ge nanostructures formation can still be studied. For instance, it would be both fascinating and beneficial to study the real time measurements of phase separation in Au-Ge system. Better understanding of growth of Au-Ge nanostructures *via* phase separation could be achieved by performing more *in-situ* experiments. This observation of phase separation at nanoscale would be very useful for proper understanding of gold contacts on Si-Ge based devices.

**4. Conclusions**

We have reported the out of plane growth of epitaxial Au-Ge nanostructures by using molecular beam epitaxy. The interesting part of this work is to get the growth parameters in UHV condition to enhance the in plane material transport which leads to formation of out of plane Au-Ge nanostructures. The effect of substrate temperature, on the formation of above structures was studied in detail.

**Figure captions:**

**Fig 1:** For as deposited MBE sample (sample A), (a) SEM micrograph showing gold nanostructures with typical size of about 27 nm (b) XTEM image showing the interface of Au thin film and Ge substrate and corresponding high resolution image (inset Fig.) shows d-spacing of Au(200) and Ge(111).

**FIG 2:** (a) SEM image taken at 54º tilt for the sample that was annealed at 500ºC in UHV chamber (sample B) (b) corresponding bright field XTEM image of the islands and (c) high resolution XTEM image of one of the nanostructure (sample B) showing inter-diffusion of Au into Ge substrate.

**FIG 3:** (a) SEM micrograph taken at RT after 2.0 nm Ge deposited on Au patterned substrate at 500ºC (sample C), (b) corresponding XTEM image shows the Au -Ge nanostructures along with the Ge nano islands (the High resolution of the square marked region is shown in the inset), (c) HRXTEM image of one of the Au-Ge islands depicts the pedestal Ge underneath the Au and (d) filtered image of the square marked region shows the formation of Au-Ge composition at the interface.

**FIG 4:** (a) SEM micrograph taken at RT after 2.0 nm Ge deposited on Au patterned substrate at 600ºC (sample D), (b) corresponding XTEM image shows the Au-Ge nanostructures, (c) HRXTEM image of one of the Au-Ge islands depicts the pedestal Ge underneath the Au and (d) filtered image of the square marked region shows the formation of Au-Ge composition at the interface.

**FIG 5:** High resolution XTEM of the sample taken at RT after 2nm Ge was deposited on Ge (100) followed by UHV annealing at 500ºC (sample E), shows the homo epitaxial growth of Ge layer on Ge (100)

**FIG 1 : Rath et al**

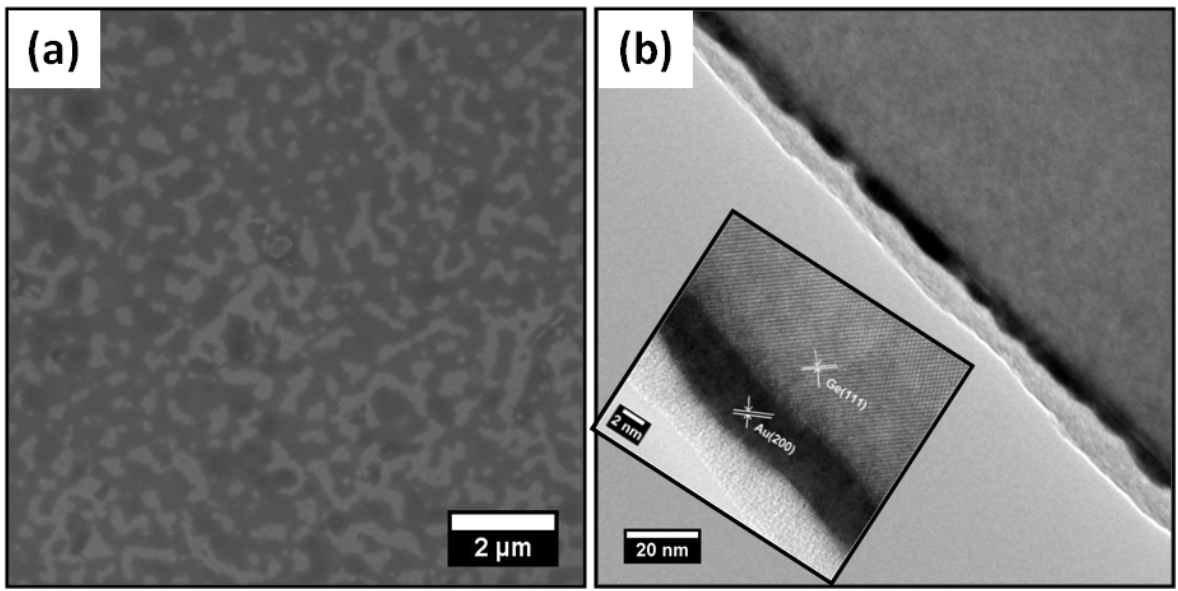

**FIG 2 : Rath et al**

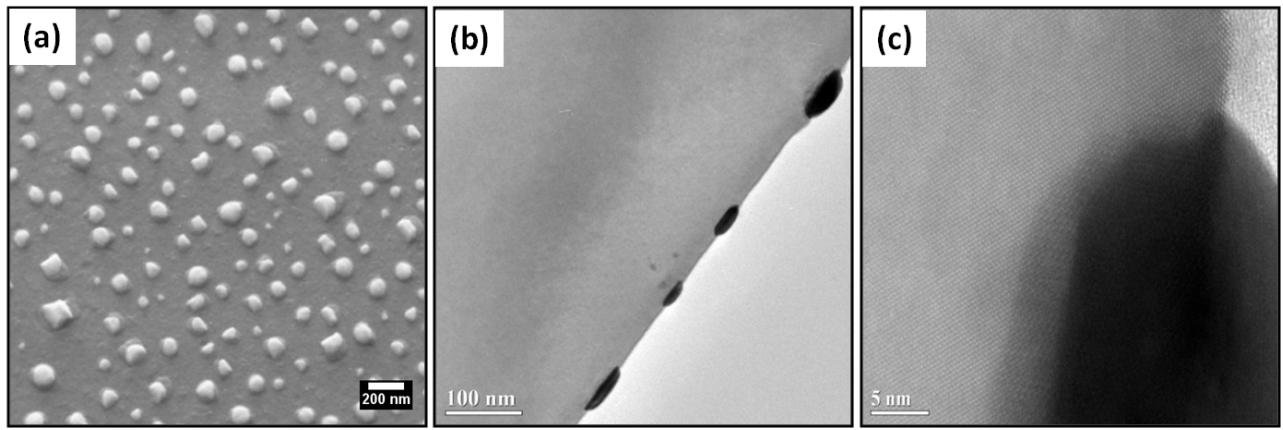

**FIG 3 : Rath et al**

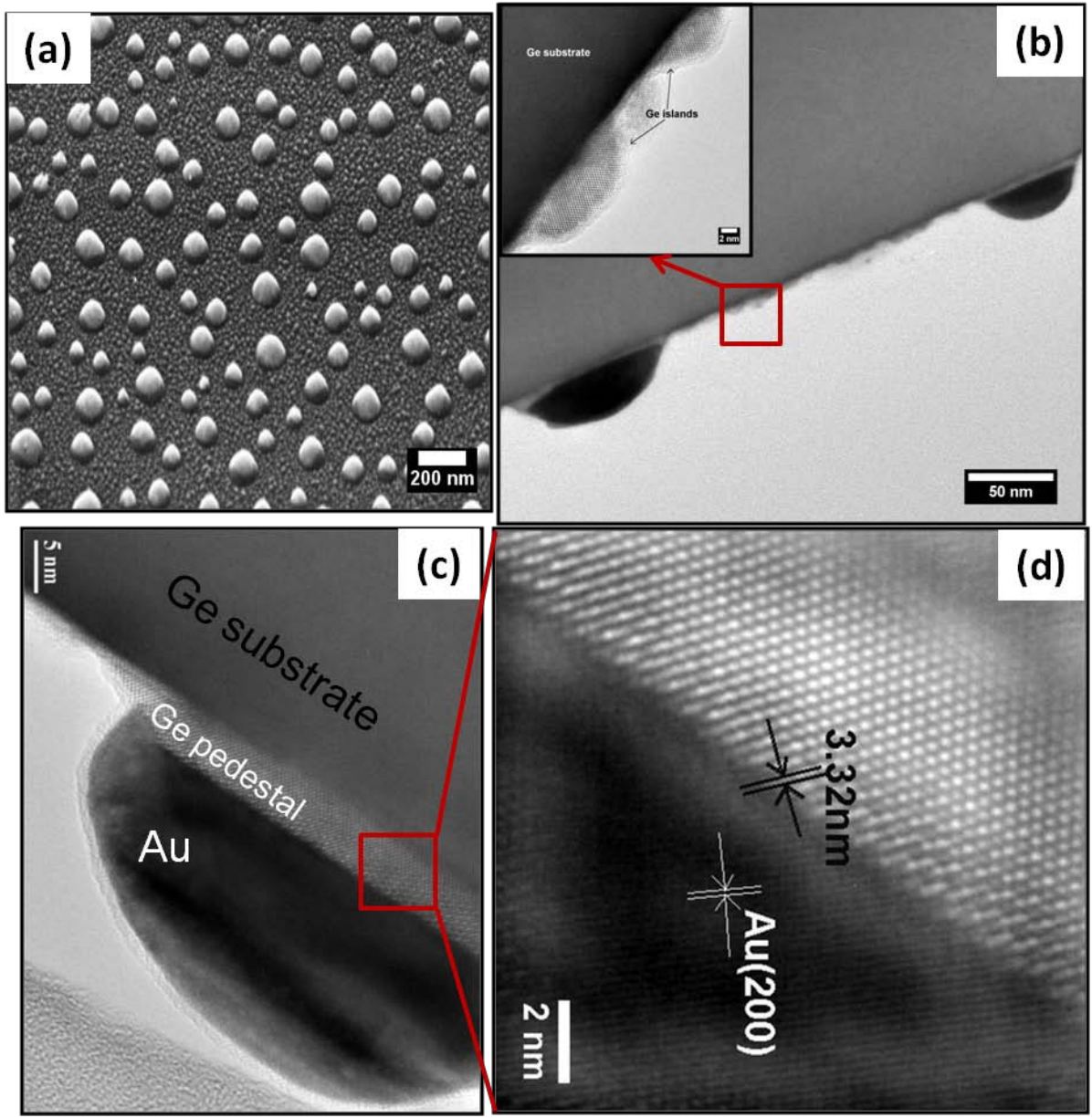

**FIG 4 : Rath et al**

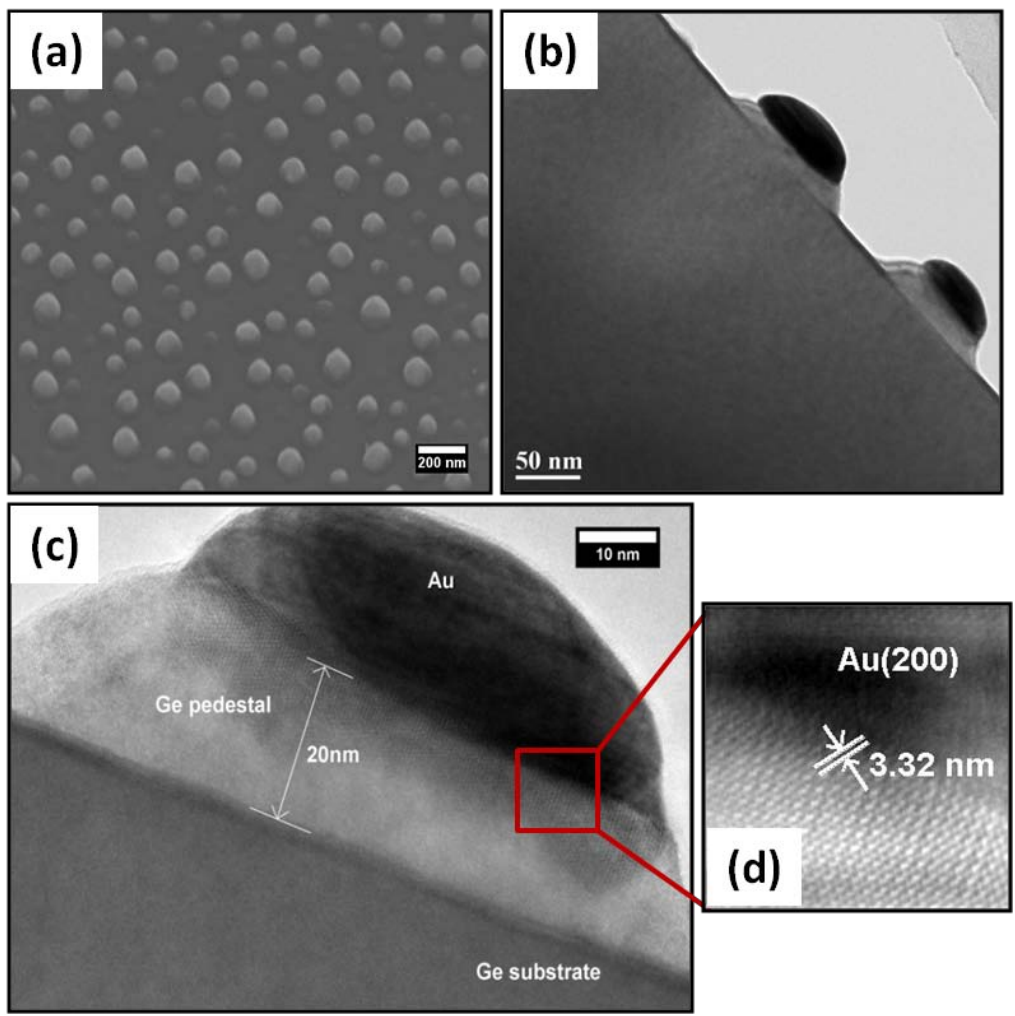

**FIG 5 : Rath et al**

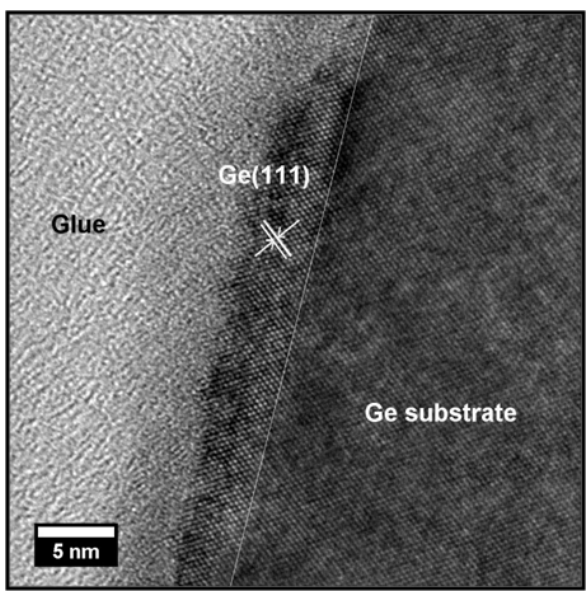